\documentclass[english,manuscript]{aastex}
\usepackage[T1]{fontenc}
\usepackage[latin9]{inputenc}
\makeatletter

\providecommand{\tabularnewline}{\\}

\makeatletter

\newcommand{\ms}{$\rm{molecules}\cdot\rm{s}^{-1}$ }
\def\las{\mathrel{\hbox{\rlap{\hbox{\lower3pt\hbox{$\sim$}}}\hbox{\raise2pt\hbox{$<$}}}}}
\def\gas{\mathrel{\hbox{\rlap{\hbox{\lower3pt\hbox{$\sim$}}}\hbox{\raise2pt\hbox{$>$}}}}}
\def\dg{{}^\g\setbox\help=\hbox{${}^\g$}\setbox\punt=\hbox{$.$}\skip
\terug=-.5\wd\help plus0pt minus0pt\advance\skip\terug by -0.17em plus0em
minus0em\hskip\skip\terug\skip\vooruit=-\skip\terug\advance\skip\vooruit by
-\wd\punt.\hskip\skip\vooruit}

\shorttitle{Phaethon}

\makeatother

\usepackage{babel}
\makeatother
\begin{document}

\title{An Upper Limit on Gas Production from 3200 Phaethon}

\author{Paul A. Wiegert\footnote{Corresponding author: P. A. Wiegert, Department of Physics and Astronomy,
The University of Western Ontario, London, Ontario, Canada N6A 3K7;
e-mail: pwiegert@uwo.ca; Phone/fax: (519) 661-2111 x81327/(519) 661-2033.}
 ~and Martin Houde}

\affil{Department of Physics and Astronomy, The University of Western Ontario,
London, Canada N6A 3K7}

\author{and Ruisheng Peng}

\affil{Caltech Submillimeter Observatory, Hilo, HI 96720, USA}

\received{01 Aug 2007}
\accepted{19 Dec 2007}
\slugcomment{Accepted for publication in Icarus}

\begin{abstract}
Asteroid 3200 Phaethon resembles a comet in some ways, including a
highly-eccentric orbit ($e\sim0.89$) and a strong associated meteor
shower (the Geminids). Yet this object has never been observed to
exhibit any cometary activity, i.e., gas production. We observed 3200
Phaethon with the Caltech Submillimeter Observatory on two occasions,
once while it was near its closest approach to Earth as it neared
perihelion, and another while it was further from Earth post-perihelion.
Observations of the $J=2\rightarrow1$ and $J=3\rightarrow2$ rotational
transitions of $^{12}$CO, typically strong lines in comets and indicative
of gas production, yielded no detection. Upper limits on the $^{12}$CO
production of $1.8\times10^{28}$ molecules~s$^{-1}$ and $7.6\times10^{28}$
molecules~s$^{-1}$ for Phaethon were determined on these two occasions. 
\end{abstract}


\section{Introduction}

Asteroid 3200 Phaethon was discovered by the Infrared Astronomical
Satellite in 1983 \citep{grekow83}. Almost immediately thereafter its
orbit was noted to be very similar to that of the Geminid meteor
shower \citep{whi83}. This makes it the only asteroid linked to a
strong meteor shower.  A possible exception is asteroid 2003~EH1,
which is associated with the Quadrantid shower
\citep{jen04,wilryabat04,wiebro05}) but which has not yet been
observed near perihelion and thus whether it is asteroidal or cometary
in nature remains unclear.

The link between Phaethon and the Geminid meteor shower is also based
on the closeness of their orbits, thus one could argue that they are
simply aligned by coincidence. However, based on the very small
difference between their orbits (see Table~1) and the absence of other
objects on orbits nearby, \citet{wiebro04a} determined that
probability of the orbital similarity of Phaethon and the Geminids
being due to a chance alignment is less than one in a thousand. Though
Phaethon's current activity (if any) is certainly low, intermittent or both,
a study of the orbits of observed Geminids has indicated that they are
consistent with cometary release no more than 2000 years ago and
possibly within the last 600 years \citep{gus89}. Thus ongoing,
perhaps sporadic, gas production by this object remains a
possibility.

If 3200~Phaethon is not a comet, this does not mean it cannot be the
parent body of the Geminids. A collision or impact might have released
the material which formed the Geminid stream. A new rare population of
{}``main-belt comets'' has been uncovered in the last several years
\citep{hsijew06}, bodies whose gas and dust production may be partly
the result of collisions. However, the sublimation of volatiles from the
body's surface also has a role to play, as evidenced by the seasonal cycle of
activity that has been seen in one of them (133P/Elst-Pizarro) to date
\cite[]{hsijewlow07}.

We note that Geminid meteors are much more durable than
typical cometary meteors: studies of their ablation in Earth's atmosphere
have shown that they have mean densities of 2.9~g~cm$^{-3}$ \citep{bab02},
the highest of any of the streams measured. This may imply that the
Geminids are made of relatively strong rocky asteroidal material rather
than more porous and fragile cometary material. However, the question
is complicated by the small (0.14~AU \citep{coo73}) perihelion distance
of the Geminid stream. Smaller than is typical of comets, this low
perihelion distance could result in extreme baking and sintering of
the meteoroid particles, producing more durable meteoroids than might
otherwise be expected of a cometary meteoroid stream.

Phaethon's association with a strong meteor shower and its highly
eccentric orbit ($e\approx0.89$) make it a candidate cometary object.
Yet in the over two decades since its discovery (which include many
perihelion passages of its 1.43-year orbit), no cometary activity has
ever been observed \citep{cocbar84,chamcfsch96,hsijew05}. 

Phaethon's spectral type is B \cite[]{gremeadav85}, a primitive type
associated with the outer portion of the main asteroid belt, though
its albedo (0.11-0.17) is somewhat higher than would be expected of
such a type \cite[]{veekowmat84,gremeadav85}, as was pointed out by
\cite{weibotlev02}.  Visible and near IR spectroscopy of Phaethon
support its identification as an asteroid rather than a comet nucleus
\cite[]{liccammot07}.  Phaethon has been also linked with the smaller
asteroid 2005~UD, with which it shares a bluish colour, an absence of
dust production \cite[]{jewhsi06} and an orbital similarity
\cite[]{ohtsekkin06}.  Is Phaethon then an asteroid, and if so, how
did it produce its meteor stream?

The question of the nature of this unusual object can only be answered
by ongoing observations. In particular, observations which might reveal
gas production by this body are the most important to undertake, as
a positive detection would unequivocally identify this object as a
comet, though its muted activity might imply it was extinct or dormant.
Given the lack of previous detection of gas from Phaethon, its gas
production rate is undoubtedly low, and observations should ideally
be conducted near perihelion, or as close to it as can be managed
given the interfering presence of the Sun. 

In this paper we describe our attempts to measure gas production on
Phaethon through the detection of $J=2\rightarrow1$ and
$J=3\rightarrow2$ rotational transitions of the $^{12}$CO molecule at
the Caltech Submillimeter Observatory (CSO), located atop Mauna Kea,
Hawaii.  Observations in the mm/sub-mm range suffer the constraints of
bad weather even more than traditional optical astronomy. This project
arose from an idea to use time on the CSO that was unsuitable for
projects requiring excellent atmospheric conditions for other
purposes.  The choice of Phaethon as a target, despite its lack of
observed activity, was motivated by its meteor stream and the
intermittency of activity seen in other comets eg. in the main-belt:
ongoing monitoring of this object seems justifiable.  The
$J=2\rightarrow1$ and $J=3\rightarrow2$ rotational transitions of
carbon monoxide (CO) are strong in bright comets, though not perhaps
ideal indicators of gas production for Phaethon. They were chosen as
being among our best bets given our observational window during
mediocre conditions ($\tau_{225~GHz} \sim 0.15$). In fact our upper
limit on gas production using these lines is comparable to that of
a moderate-activity comet, and by observing as close to perihelion as
possible, we hoped to catch Phaethon in outburst. Though our attempt
was unsuccessful, only by ongoing observations will Phaethon's nature
be clarified.

We present the three spectra we obtained from corresponding sets of
observations and discuss our results in the next sections.

\section{Observations}

The three sets of observations were taken at the Caltech Submillimeter
Observatory, with the 200-300 GHz receiver on 5 November 2004 and 13
September 2005 for the $^{12}$CO($J=2\rightarrow1$) transition, and
with the 300-400 GHz receiver on 5 November 2004 for the
$^{12}$CO($J=3\rightarrow2$) transition. All spectra were obtained
with the high-resolution acousto-optics spectrometer (AOS) of 50 MHz
total bandwidth and approximately 100 kHz resolution. The CSO has a
10.4-meter dish and a beam diameter of approximately 30 and 20
arcseconds at the frequencies in question (i.e., approximately 230 GHz
and 345 GHz for $^{12}$CO $J=2\rightarrow1$ and $J=3\rightarrow2$,
respectively). For all of our observations, the data were taken using
the beam switching mode (i.e., by wobbling the secondary mirror) to
allow for cancellation of sky background emission. While reducing the
data we have set the receiver efficiencies to 70\% for both
transitions.  This corresponds to situations where not only the main
beam of the telescope probes the source, but also the sidelobes. This
technique has been used by other investigators observing
cometary lines at the CSO (eg. \cite{bocliswin00}). We assume that this
would be applicable to our observations if there were a gas cloud
surrounding 3200 Phaethon from which the sought transitions were
detected (see Discussion below) .

\begin{figure}
\epsscale{0.8}\plotone{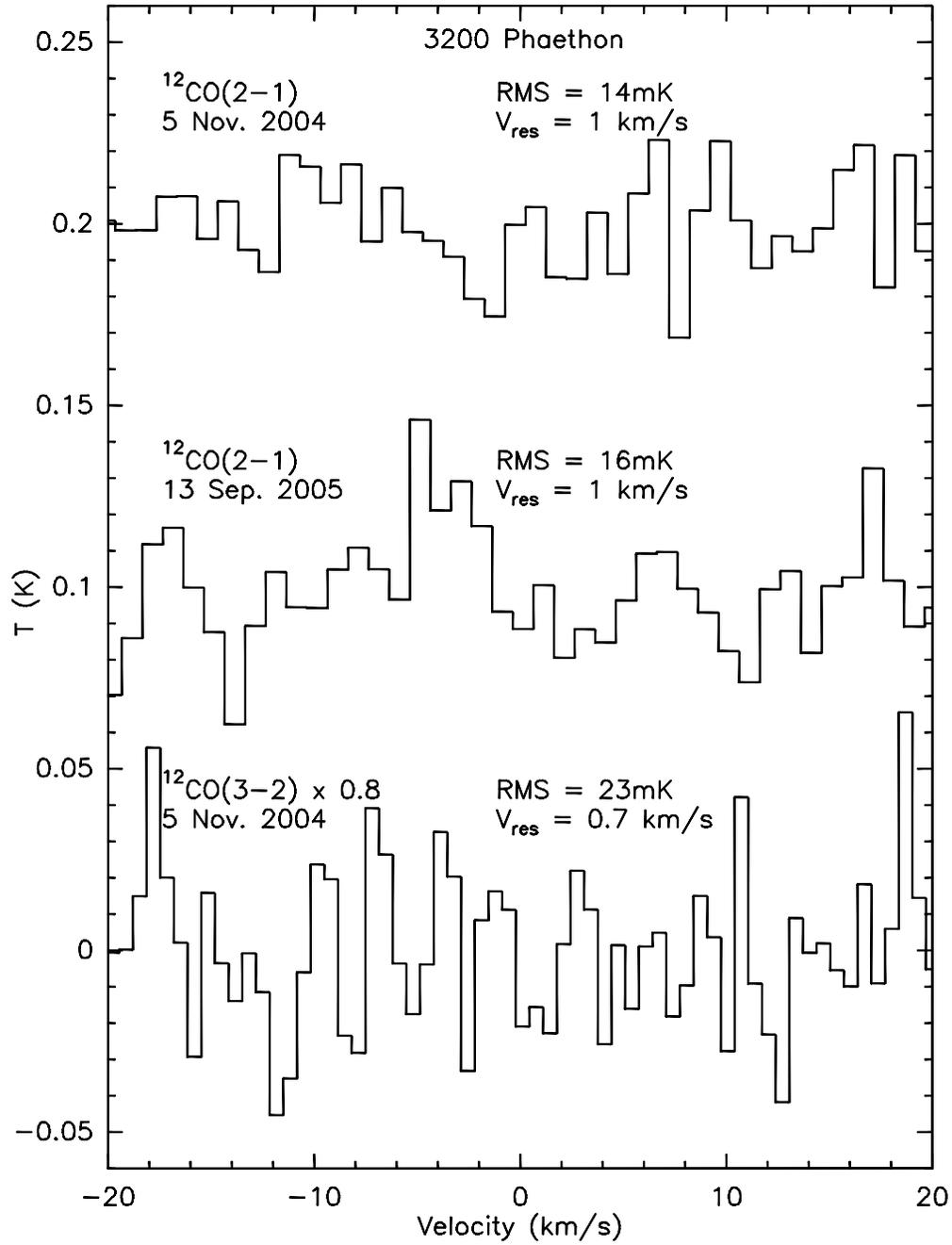}

\caption{\label{fig:spectra}The three spectra that resulted from our observations.
The top two are for our attempts at a detection of the $J=2\rightarrow1$
transition of $^{12}$CO, while the the bottom spectrum corresponds
to our $J=3\rightarrow2$ observations. }
\end{figure}

For the observations obtained on 5 November 2004, 3200 Phaethon was
1.05~AU from Earth, allowing us a close look at the object as it
passed, and 1.894~AU from the Sun, approaching perihelion. Its
elevation angle varied from $54^{\circ}$ to $28^{\circ}$ in the
morning sky, as it was moving rapidly towards the Sun. We obtained 50
minutes of observation (ON-source), but the $J=2\rightarrow1$
rotational transition line from $^{12}$CO was not detected. Given
mediocre sky conditions ($\tau_{225\,{\rm {GHz}}}\simeq0.15$) and a
system temperature of $T_{{\rm {sys}}}\simeq444{\rm {K}}$, we were
able to integrate down to a noise floor of 14~mK (with a velocity
resolution of $\simeq1\,{\rm {km\, s^{-1}}}$ after {}``smoothing'' the
spectrum to lower the noise level), and determine an upper limit
($3\sigma$) of approximately $T\simeq42\,{\rm {mK}}$ for the
brightness temperature of the line. We also attempted to detect the
$J=3\rightarrow2$ transition from $^{12}$CO previously on the same
night, when Phaethon was at an elevation of $70^{\circ}$ to
$59^{\circ}$ with a similar outcome. For these observations, with a
system temperature $T_{{\rm {sys}}}\simeq918{\rm {K}}$, the 48 minutes
of ON-source integration yielded an upper limit of
$T\simeq69\,{\rm {mK}}$ in a velocity resolution of
$\simeq0.7\,{\rm {km\, s^{-1}}}$. Our last set of data was obtained on
13 September 2005 when we again attempted to observe the
$J=2\rightarrow1$ transition from $^{12}$CO under comparable observing
conditions (i.e., $\tau_{225\,{\rm {GHz}}}\simeq0.16$ and $T_{{\rm
    {sys}}}\simeq440{\rm {K}}$). On that night, 3200 Phaethon was
1.904~AU from Earth and 2.298~AU from the Sun approaching aphelion.
Earlier observations nearer perihelion were impossible because Phaethon
had been in the daytime sky for many months.  This new set of data (50
minutes of integration ON-source), obtained at elevation angles
ranging from $35^{\circ}$ to $70^{\circ}$, yielded no detection and an
upper limit of approximately $T\simeq48\,{\rm {mK}}$, again with
a velocity resolution of $\simeq1\,{\rm {km\, s^{-1}}}$.  The spectra
that resulted from our observations are presented in Figure
\ref{fig:spectra}.

\section{Discussion}

The submillimeter rotational transitions of $^{12}$CO have been studied
in detail for comets and the strength of the lines can be linked to
gas production rates \citep{croleb83}.

Assuming a lifetime of $^{12}$CO against dissociation by solar UV
radiation at 1~AU of $\tau=1.5\times10^{6}$~s (\citet{huecar79} from
\cite{croleb83}) and assuming a gas expansion velocity $v_{{\rm {\rm
}{exp}}}=800$ m/s (the pre-perihelion value determined for Hyakutake
from submillimeter observations by \citet{bivboccro99}), any $^{12}$CO
cloud around 3200 Phaethon should have a diameter
$\sim2.4\times10^{6}$~km, with an angular size of about 0.9 degree at
1~AU. A comparison of this angular size with the CSO main beam sizes
quoted earlier for our observations justifies the efficiency adopted
for our observations (i.e., 70\%).

Assuming fluorescence equilibrium and a Haser distribution \citep{has57,bercosque98}
with $\gamma_{{\rm {\rm }{CO}}}=v_{{\rm {\rm }{exp}}}\tau=1.2\times10^{6}$~km,
and using an emission constant $T_{B}\Delta v/\langle N_{{\rm {\rm }{CO}}}\rangle$
of $3.65\times10^{-16}$~K~km~s$^{-1}$~(molec. cm$^{-2}$)$^{-1}$
and $8.08\times10^{-16}$~K~km~s$^{-1}$~(molec. cm$^{-2}$)$^{-1}$
for the $J=2\rightarrow1$ and $J=3\rightarrow2$ transitions, respectively
\citep{croleb83}, we can derive a corresponding upper limit on $Q_{{\rm {\rm {\rm }{CO}}}}$,
the rate of $^{12}$CO production, of approximately $1.8\times10^{28}$~\ms
for 5 November 2004 (from the $J=3\rightarrow2$ transition). More
precisely, at 1.05~AU (the distance between Earth and 3200 Phaethon
during the corresponding observations) the dependence on the measured
and assumed parameters can be expressed as

\begin{equation}
Q_{{\rm {\rm }{CO}}}\approx3.60\times10^{28}\left(\frac{v_{{\rm {\rm }{exp}}}}{800~\textrm{{m/s}}}\right)^{2}\left(\frac{T_{B}}{42~\textrm{{mK}}}\right)\textrm{molecules}\cdot\textrm{s}^{-1}\label{eq:Q2-1}\end{equation}

\noindent for the $J=2\rightarrow1$ transition, and

\begin{equation}
Q_{{\rm {\rm }{CO}}}\approx1.77\times10^{28}\left(\frac{v_{{\rm {\rm }{exp}}}}{800~\textrm{{m/s}}}\right)^{2}\left(\frac{T_{B}}{69~\textrm{{mK}}}\right)\textrm{molecules}\cdot\textrm{s}^{-1}\label{eq:Q3-2}\end{equation}

\noindent for the $J=3\rightarrow2$ transition. Using an equation
similar to equation (\ref{eq:Q2-1}) for a distance of 1.904~AU we also
establish an upper limit of $Q_{{\rm {\rm
}{CO}}}\approx7.6\times10^{28}$~\ms for 13 September 2005.

These upper limits are lower than production rates observed for very
bright comets at the same heliocentric distance, e.g., C/1995 O1
(Hale-Bopp), $Q_{{\rm {\rm }{CO}}}=2.07\times10^{30}$~\ms
\citep{dismumdel01}. The limits are comparable to the production rate of
C/1996 B2 Hyakutake, $Q_{{\rm {\rm }{CO}}}\approx5\times10^{28}$~\ms
near 1~AU \citep{bivboccro99}, though it is a long-period comet and
expected to be rich in ices. 

Comet 2P/Encke, an old low-activity comet, might provide a better
benchmark. Though its carbon monoxide production has not been measured
to our knowledge, we can estimate this from its H$_{2}$O production
rate, which has measured at 0.8 to $2.6\times10^{28}$
molecules~s$^{-1}$ at similar heliocentric distances
\citep{maksilsch01}. The $^{12}$CO to H$_{2}$O ratio in comets is
variable, ranging from 0.5\% to 22\% \citep{bivboccro99,
dismumdel01,mumdeldis01b, mumdeldis01, mummcldis01,
disdelmag02,felweabur02,dismumdel03,disandvil07}. Taking the ratio to
be 5\%, we can derive a corresponding expected $^{12}$CO production
rate of $0.4\times10^{27}$ to $1.3\times10^{27}$~molecules~s$^{-1}$
for comet Encke. Though a nearly-extinct comet could produce $^{12}$CO
at even lower rates, this presents us with a target measurement
sensitivity of $Q_{\mathrm{CO}}\sim10^{27}$~molecules~s$^{-1}$.

In view of these numbers it will be necessary to make deeper observations
of 3200 Phaethon in order to detect any $^{12}$CO production or to
reduce significantly our derived upper limits.  It follows that we
need to increase our sensitivity by at least an order of
magnitude. The results discussed here were obtained under relatively
poor sky conditions and improving these limits should be feasible with
a combination of longer exposure times and better observing conditions
(e.g., $\tau_{225\,{\rm {GHz}}}\simeq0.05$), as is relatively common
(20\% to 25\% of the time) on Mauna Kea. The sensitivity can also be
increased by observing Phaethon when it is closest to Earth, as the
detection limit drops approximately linearly with the Earth-comet
distance. Observations of this object during its upcoming close
approach to Earth ($\las0.13$~AU in December 2007) would thus be
particularly recommended.

\acknowledgements{M.H.'s research is funded through the NSERC
Discovery Grant, Canada Research Chair, Canada Foundation for
Innovation, Ontario Innovation Trust, and the University of Western
Ontario's Academic Development Fund programs. This work was supported
in part by the Natural Sciences and Engineering Research Council of
Canada. The Caltech Submillimeter Observatory is funded by the NSF
through contract AST 05-40882.}

\newpage{}\bibliographystyle{icarusbib} 
\bibliography{Wiegert}

\newpage{}

\clearpage{}
\begin{table}
\centerline{\begin{tabular}{lccccccc}
\hline 
 & $a$ (AU)  & $e$  & $q$ (AU)  & $i$ (deg)  & $\Omega$ (deg) & $\omega$ (deg) & $T_{J}$\tabularnewline
\hline 
3200 Phaethon & 1.27 & 0.890 & 0.140 & 22.2 & 265.4 & 322.0 & 4.51 \tabularnewline
Geminids      & 1.36 & 0.896 & 0.142 & 23.6 & 261.0 & 324.3 & 4.23 \tabularnewline
\hline
\end{tabular}}

\caption{Standard orbital elements of 3200 Phaethon (NeoDys website http://newton.dm.unipi.it/cgi-bin/neodys/neoibo) and the Geminid meteor shower \citep{coo73}.}
\end{table}

\end{document}